# Tau Physics with Polarized Beams*

Mourad Daoudi

*Stanford Linear Accelerator Center*
*Stanford, CA 94309*

Representing the SLD Collaboration

**Abstract**

We present the first results on tau physics using polarized beams. These include measurements of the $\tau$ Michel parameters $\xi$ and $\xi\delta$ and the $\tau$ neutrino helicity $h_\nu$. The measurements were performed using the SLD detector at the Stanford Linear Collider (SLC).

*Work supported by U.S. Department of Energy contract DE-AC03-76SF00515.

# 1 Introduction

In the reaction $e^+e^- \to Z^0 \to \tau^+\tau^-$, final-state $\tau$ polarization arises due to the asymmetric couplings of the $Z^0$ to the electron and the $\tau$. With the addition of electron beam polarization the degree of $\tau$ polarization is greatly enhanced. This is illustrated by:

$$P_\tau(\cos\theta, P_e) = -\frac{A_\tau + 2\,\frac{A_e - P_e}{1 - A_e\ P_e}\,\frac{\cos\theta}{1+\cos^2\theta}}{1 + 2\ A_\tau\,\frac{A_e - P_e}{1 - A_e\ P_e}\,\frac{\cos\theta}{1+\cos^2\theta}}, \qquad (1)$$

where $P_e$ is the average $e^-$ beam polarization and $\theta$ the $\tau$ production angle. By measuring $P_e$ and $\theta$, we determine $P_\tau$ on an event-by-event basis. The dependence on $A_e$ and $A_\tau$ is very small for large values of $P_e$.

At SLD, we exploit these features to gain in experimental sensitivity and perform measurements even with limited statistics. The double-differential cross section for the production and decay of the $\tau$ can be expressed as a function of two terms, one depending entirely on the final-state kinematic variables, and one proportional to $P_\tau(\cos\theta, P_e)$. For the two-body decay $\tau \to \pi\nu$, we have

$$\frac{d^2\sigma}{d\cos\theta dx} = 1 + h_\nu P_\tau(2x - 1), \qquad (2)$$

where $h_\nu$ is the $\tau$ neutrino helicity. In the case of the leptonic decays $\tau \to l\bar{\nu}\nu$ ($l = e, \mu$), following the Michel parametrization[1], we have

$$\frac{d^2\sigma}{d\cos\theta dx} = [f_1(x) + \rho f_2(x)] - \xi P_\tau[g_1(x) + \delta g_2(x)]. \qquad (3)$$

In these equations, $x$ is the scaled final-state pion or lepton energy, and $f_1$, $f_2$, $g_1$, and $g_2$ are simple polynomial functions. Using an unbinned maximum likelihood fit, we perform a direct measurement of the parameters associated with the $\tau$ polarization, namely $h_\nu$, $\xi$, and $\xi\delta$. In other measurements[2], the extraction of these parameters is possible only by exploiting the spin correlation between the two $\tau$'s in the event. In our case, with the added knowledge of $P_\tau$, this is not necessary, and we make use of every identified decay in our measurement.

# 2 Final State Selection and Measurement

The events used in this analysis were selected from a data sample of $150k$ hadronic $Z^0$'s collected by the SLD detector at the SLC ($50k$ with $P_e = 63.1 \pm 1.1$ and $100k$ with $P_e = 77.3 \pm 0.6$). A detailed description of the $\tau$ sample selection can be found in Ref. 3. A total of 4500 $\tau$-pair candidates have been selected. Particle identification requirements based on the combination of tracking, calorimetry, and muon detection information are applied in order to select the final states used in the measurement. The number of selected decays is given in Table 1, along with the particle identification efficiency and purity for each species. The background contamination in the pion sample is dominated by feed down from the $\rho$ and $a_1$ decay channels. In all three final-state samples, the background from Bhabha scattering and $\mu$-pair events is estimated at $\sim 1\%$. Currently, the analysis is limited to the barrel region of the detector ($\cos\theta \le 0.74$).

Table 1: Number of selected decays and particle identification efficiency and purity.

| Channel | # Decays | Efficiency (%) | Purity (%) |
|---|---|---|---|
| $\tau \rightarrow \pi\nu$ | 414 | 48 | 84 |
| $\tau \rightarrow e\bar{\nu}\nu$ | 863 | 80 | 98 |
| $\tau \rightarrow \mu\bar{\nu}\nu$ | 1137 | 92 | 94 |

As a demonstration of the power of this analysis, we plot the final-state pion and lepton energy spectra in Figs. 1 and 2. In Fig. 1 the dots represent the data and the histograms Monte Carlo. Fig. 1a contains the pion spectrum for the following two combinations:

1. Left-handed beam polarization and the $\tau^-$ ($\tau^+$) emitted in the forward (backward) region of the detector.
2. Right-handed beam polarization and the $\tau^-$ ($\tau^+$) emitted in the backward (forward) region of the detector.

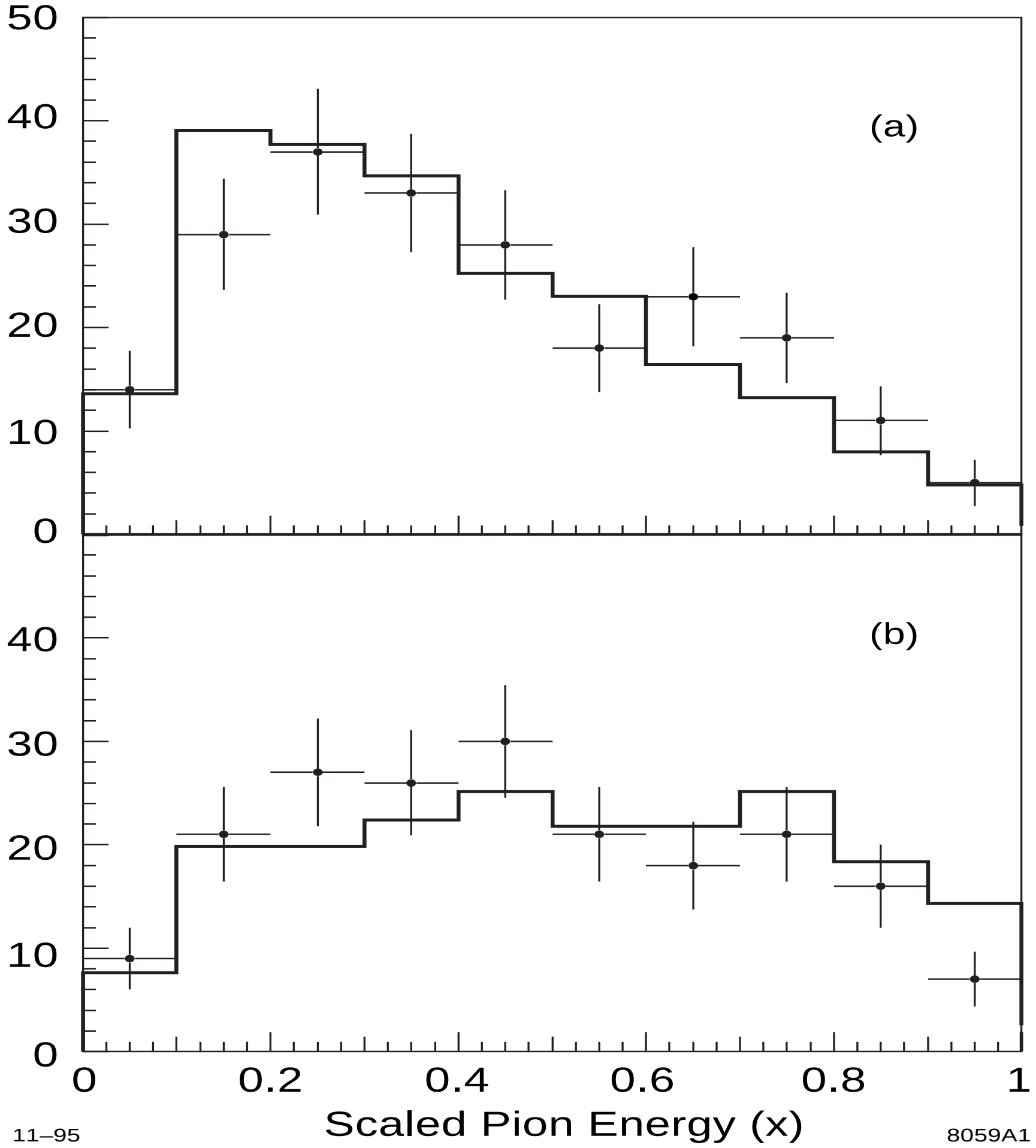

Figure 1: Scaled pion energy in $\tau \rightarrow \pi\nu$ decays.

Here, the $\tau^-$ ($\tau^+$) is predominantly left-handed (right-handed), and the pion spectrum is soft as expected for a two-body decay. On the other hand, for the two opposite combinations of $P_e$ and $\theta$ in Fig. 1b, the spectrum is hard since the pion comes from the decay of a predominantly right-handed (left-handed) $\tau^-$ ($\tau^+$). The clear distinction between the two spectra is a powerful indication of how the presence of beam polarization allows one to infer the helicity of the $\tau$ in the data.

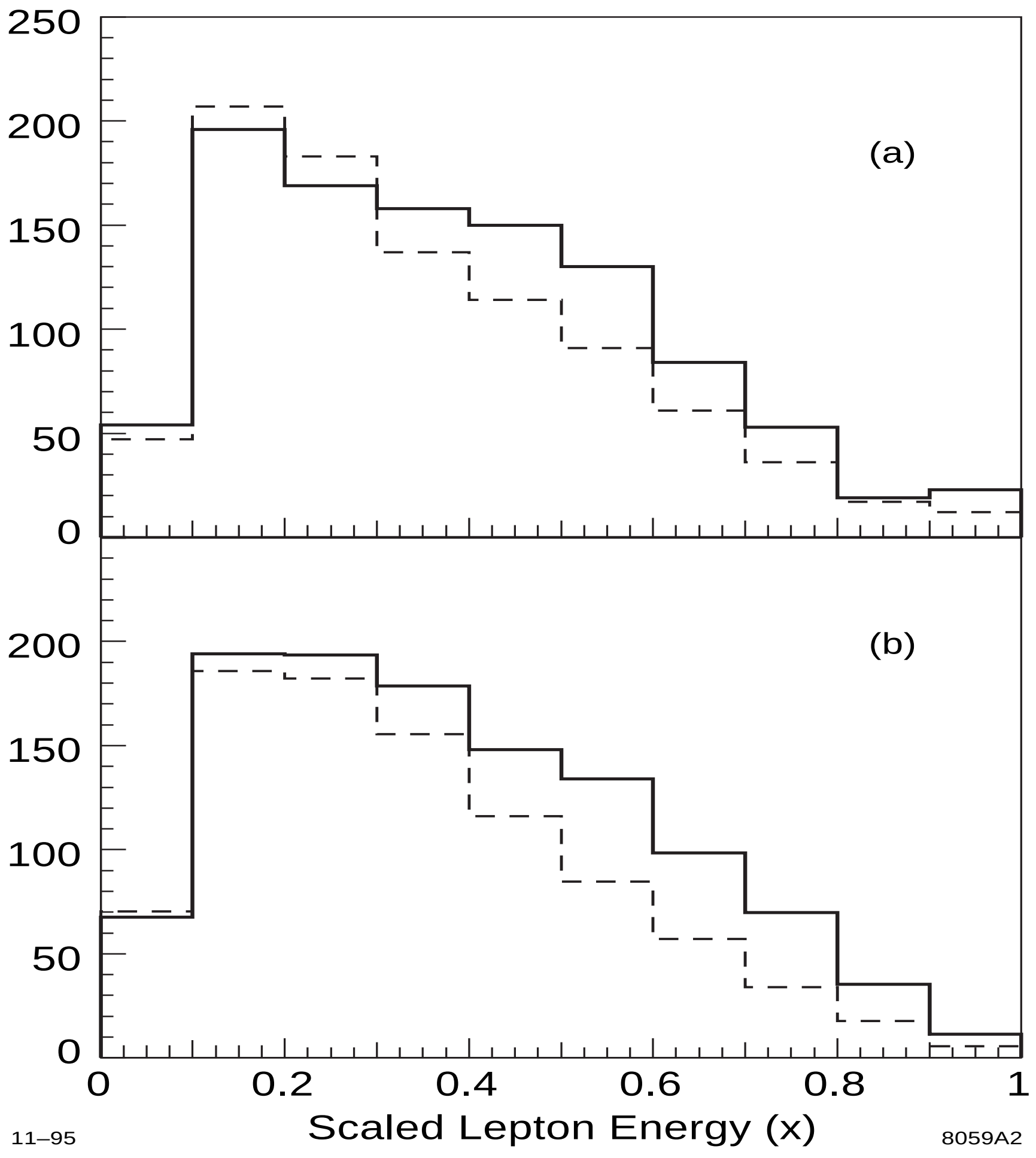

Figure 2: Scaled lepton energy in $\tau \to l\bar{\nu}\nu$ decays.

For the $\tau \to e\bar{\nu}\nu$ and $\tau \to \mu\bar{\nu}\nu$ decays, the lepton energy spectrum is represented in Fig. 2 by the solid and dashed histograms for the above two combinations of $P_e$ and $\theta$, respectively. The data are plotted in Fig. 2a and the Monte Carlo in Fig. 2b. Again, a clear distinction is seen between the two combinations, though not as pronounced due to the three-body nature of the decays.

We perform an unbinned maximum likelihood fit to the theoretical analytic functions of Eqs. 2 and 3. Several corrections[4] are applied, including selection efficiency, detector resolution, background, and radiative effects. These are parameterized as a function of the scaled energy $x$ and the $\tau$ production angle $\theta$ using Monte Carlo[5]. In the analysis of the lepton sample, the parameter $\rho$ is kept fixed, and we fit for the parameters $\xi$ and $\xi\delta$. In addition, lepton universality is assumed and the electron and muon samples are combined. We use a value of 0.15 for both $A_e$ and $A_\tau$, which is approximately the average of the SLC and LEP results[6].

## 3 Results and Conclusions

We obtain the following results:

$$\begin{aligned} h_\nu &= -0.89 \pm 0.21\ (stat) \pm 0.07\ (syst), \\ \xi &= \ \ 1.17 \pm 0.35\ (stat) \pm 0.21\ (syst), \\ \xi\delta &= \ \ 0.49 \pm 0.24\ (stat) \pm 0.13\ (syst). \end{aligned} \tag{4}$$

These results are consistent with present world averages and with the $(V - A)$ nature of the charged weak current predicted by the Standard Model. They are preliminary and the systematic errors assigned at this stage of the analysis are conservative and expected to be reduced. The main systematic error for $h_\nu$ is the uncertainty in the $\pi\nu$ final state selection efficiency and purity. For $\xi$ and $\xi\delta$, it is dominated by the detector resolution and radiative effects. We expect a substantial increase in our data sample with high beam polarization. Furthermore, the installation of a new CCD pixel vertex detector will improve our polar angle coverage to $\cos\theta \approx 0.85$.

With these improvements, we anticipate making a useful contribution in the field of $\tau$ physics with polarized beams, which we have shown represents a unique environment for probing the decay of the $\tau$ lepton.

## Acknowledgments

I wish to thank my colleagues Nik Allen, Erez Etzion, Jim Johnson, and Jim Quigley for their valuable input.

## The SLD Collaboration

K. Abe,[29] I. Abt,[14] C.J. Ahn,[26] T. Akagi,[27] N.J. Allen,[4] W.W. Ash,[27]† D. Aston,[27] K.G. Baird,[25] C. Baltay,[33] H.R. Band,[32] M.B. Barakat,[33] G. Baranko,[10] O. Bardon,[16] T. Barklow,[27] A.O. Bazarko,[11] R. Ben-David,[33] A.C. Benvenuti,[2] T. Bienz,[27] G.M. Bilei,[22] D. Bisello,[21] G. Blaylock,[7] J.R. Bogart,[27] T. Bolton,[11] G.R. Bower,[27] J.E. Brau,[20] M. Breidenbach,[27] W.M. Bugg,[28] D. Burke,[27] T.H. Burnett,[31] P.N. Burrows,[16] W. Busza,[16] A. Calcaterra,[13] D.O. Caldwell,[6] D. Calloway,[27] B. Camanzi,[12] M. Carpinelli,[23] R. Cassell,[27] R. Castaldi,[23][a] A. Castro,[21] M. Cavalli-Sforza,[7] E. Church,[31] H.O. Cohn,[28] J.A. Coller,[3] V. Cook,[31] R. Cotton,[4] R.F. Cowan,[16] D.G. Coyne,[7] A. D'Oliveira,[8] C.J.S. Damerell,[24] M. Daoudi,[27] R. De Sangro,[13] P. De Simone,[13] R. Dell'Orso,[23] M. Dima,[9] P.Y.C. Du,[28] R. Dubois,[27] B.I. Eisenstein,[14] R. Elia,[27] E. Etzion,[4] D. Falciai,[22] M.J. Fero,[16] R. Frey,[20] K. Furuno,[20] T. Gillman,[24] G. Gladding,[14] S. Gonzalez,[16] G.D. Hallewell,[27] E.L. Hart,[28] Y. Hasegawa,[29] S. Hedges,[4] S.S. Hertzbach,[17] M.D. Hildreth,[27] J. Huber,[20] M.E. Huffer,[27] E.W. Hughes,[27] H. Hwang,[20] Y. Iwasaki,[29] D.J. Jackson,[24] P. Jacques,[25] J. Jaros,[27] A.S. Johnson,[3] J.R. Johnson,[32] R.A. Johnson,[8] T. Junk,[27] R. Kajikawa,[19] M. Kalelkar,[25] H. J. Kang,[26] I. Karliner,[14] H. Kawahara,[27] H.W. Kendall,[16] Y. Kim,[26] M.E. King,[27] R. King,[27] R.R. Kofler,[17] N.M. Krishna,[10] R.S. Kroeger,[18] J.F. Labs,[27] M. Langston,[20] A. Lath,[16] J.A. Lauber,[10] D.W.G. Leith,[27] M.X. Liu,[33] X. Liu,[7] M. Loreti,[21] A. Lu,[6] H.L. Lynch,[27] J. Ma,[31] G. Mancinelli,[22] S. Manly,[33] G. Mantovani,[22] T.W. Markiewicz,[27] T. Maruyama,[27] R. Massetti,[22] H. Masuda,[27] E. Mazzucato,[12] A.K. McKemey,[4] B.T. Meadows,[8] R. Messner,[27] P.M. Mockett,[31] K.C. Moffeit,[27] B. Mours,[27] G. Müller,[27] D. Muller,[27] T. Nagamine,[27] U. Nauenberg,[10] H. Neal,[27] M. Nussbaum,[8] Y. Ohnishi,[19] L.S. Osborne,[16] R.S. Panvini,[30] H. Park,[20] T.J. Pavel,[27] I. Peruzzi,[13][b] M. Piccolo,[13] L. Piemontese,[12] E. Pieroni,[23] K.T. Pitts,[20] R.J. Plano,[25] R. Prepost,[32] C.Y. Prescott,[27] G.D. Punkar,[27] J. Quigley,[16] B.N. Ratcliff,[27] T.W. Reeves,[30] J. Reidy,[18] P.E. Rensing,[27] L.S. Rochester,[27] J.E. Rothberg,[31] P.C. Rowson,[11] J.J. Russell,[27] O.H. Saxton,[27] S.F. Schaffner,[27] T. Schalk,[7] R.H. Schindler,[27] U. Schneekloth,[16] B.A. Schumm,[15] A. Seiden,[7] S. Sen,[33] V.V. Serbo,[32] M.H. Shaevitz,[11] J.T. Shank,[3] G. Shapiro,[15] S.L. Shapiro,[27] D.J. Sherden,[27] K.D. Shmakov,[28] C. Simopoulos,[27] N.B. Sinev,[20] S.R. Smith,[27] J.A. Snyder,[33] P. Stamer,[25] H. Steiner,[15] R. Steiner,[1] M.G. Strauss,[17] D. Su,[27] F. Suekane,[29] A. Sugiyama,[19] S. Suzuki,[19] M. Swartz,[27] A. Szumilo,[31] T. Takahashi,[27] F.E. Taylor,[16] E. Torrence,[16] J.D. Turk,[33] T. Usher,[27] J. Va'vra,[27] C. Vannini,[23] E. Vella,[27] J.P. Venuti,[30] R. Verdier,[16] P.G. Verdini,[23] S.R. Wagner,[27] A.P. Waite,[27] S.J. Watts,[4] A.W. Weidemann,[28] E.R. Weiss,[31] J.S. Whitaker,[3] S.L. White,[28] F.J. Wickens,[24] D.A. Williams,[7] D.C. Williams,[16] S.H. Williams,[27] S. Willocq,[33] R.J. Wilson,[9] W.J. Wisniewski,[5] M. Woods,[27] G.B. Word,[25] J. Wyss,[21] R.K. Yamamoto,[16] J.M. Yamartino,[16] X. Yang,[20] S.J. Yellin,[6] C.C. Young,[27] H. Yuta,[29] G. Zapalac,[32] R.W. Zdarko,[27] C. Zeitlin,[20] Z. Zhang,[16] and J. Zhou,[20]

[1] *Adelphi University, Garden City, New York 11530*
[2] *INFN Sezione di Bologna, I-40126 Bologna, Italy*
[3] *Boston University, Boston, Massachusetts 02215*
[4] *Brunel University, Uxbridge, Middlesex UB8 3PH, United Kingdom*
[5] *California Institute of Technology, Pasadena, California 91125*
[6] *University of California at Santa Barbara, Santa Barbara, California 93106*
[7] *University of California at Santa Cruz, Santa Cruz, California 95064*
[8] *University of Cincinnati, Cincinnati, Ohio 45221*
[9] *Colorado State University, Fort Collins, Colorado 80523*
[10] *University of Colorado, Boulder, Colorado 80309*
[11] *Columbia University, New York, New York 10027*
[12] *INFN Sezione di Ferrara and Università di Ferrara, I-44100 Ferrara, Italy*
[13] *INFN Lab. Nazionali di Frascati, I-00044 Frascati, Italy*
[14] *University of Illinois, Urbana, Illinois 61801*
[15] *Lawrence Berkeley Laboratory, University of California, Berkeley, California 94720*
[16] *Massachusetts Institute of Technology, Cambridge, Massachusetts 02139*
[17] *University of Massachusetts, Amherst, Massachusetts 01003*

[18] *University of Mississippi, University, Mississippi 38677*
[19] *Nagoya University, Chikusa-ku, Nagoya 464 Japan*
[20] *University of Oregon, Eugene, Oregon 97403*
[21] *INFN Sezione di Padova and Università di Padova, I-35100 Padova, Italy*
[22] *INFN Sezione di Perugia and Università di Perugia, I-06100 Perugia, Italy*
[23] *INFN Sezione di Pisa and Università di Pisa, I-56100 Pisa, Italy*
[25] *Rutgers University, Piscataway, New Jersey 08855*
[24] *Rutherford Appleton Laboratory, Chilton, Didcot, Oxon OX11 0QX United Kingdom*
[26] *Sogang University, Seoul, Korea*
[27] *Stanford Linear Accelerator Center, Stanford University, Stanford, California 94309*
[28] *University of Tennessee, Knoxville, Tennessee 37996*
[29] *Tohoku University, Sendai 980 Japan*
[30] *Vanderbilt University, Nashville, Tennessee 37235*
[31] *University of Washington, Seattle, Washington 98195*
[32] *University of Wisconsin, Madison, Wisconsin 53706*
[33] *Yale University, New Haven, Connecticut 06511*
† *Deceased*
[a] *Also at the Università di Genova*
[b] *Also at the Università di Perugia*